\documentstyle[12pt]{article}

\begin{document}
\thispagestyle{empty}
\hspace{12cm}
IHEP  97-24
\hspace{10cm}
\vspace{1cm} 
\vbox{\vglue 1cm}

\begin{center}
{\Large \bf Associated Production of {\boldmath $H^0\gamma$} or 
{\boldmath $H^0 Z^0$} pairs at {\boldmath $\mu^+ \mu^-$} Collisions.}
\end{center}

\vspace{0.2cm}
\begin{center}
{\bf V.A.LITVIN}, ~~ {\bf F.F.TIKHONIN} \\
\end{center}
\vspace{0.2cm}
\begin{center}
{\bf Abstract}
\end{center}

   We calculate the cross--sections for the production of Standard Model
Higgs Boson in association with the neutral gauge bosons (photon and
Z--boson). For the case of reaction $\mu^+ \mu^- \, \to \, \gamma H^0$,
complete and compact analytical expressions for the differential and total
cross--sections,  applicable also for the case of any (pseudo)scalar 
particle with mass-coupling proportionality, in particulary, for an axion
are given.
Reaction $\mu^+ \mu^- \, \to \, Z^0 H^0$ is some 
"generalization" of the well-known "Bjorken process" for the case of
$ e^+ e^-$  collisions. Various distributions for both the processes
above are illustrated for the energies, which will be reached at future
$\mu^+ \mu^-$ colliders. Study of those processes will be evidently 
complementary to the precision measurements at the Higgs resonance region.

\vspace{3cm}
\begin{center}
{\bf Protvino, 1997}
\end{center}

\newpage

\section{Introduction}

    The search for Higgs particles from various models and the 
study of the sundry scenario for electroweak symmetry breaking 
mechanism is one of the most important goals of future 
high energy colliders \cite{1}. Another important task is to make a 
detailed study of the basic properties of possible Higgs particles in
various models. One of the most interesting and crucial parameters are 
the values of the couplings of a Higgs bosons to other fundamental
particles \cite{2}. Measurements of those couplings would allow one to make 
choice between different Higgs schemes, but in this short note we restrict 
our discussions to the Standard Model (SM) with a single neutral Higgs boson.
    
    Presumably the fundamental scalar will be revealed and 
investigated to some extent at the forthcoming LHC collider, but the
precision will be evidently insufficient for the aims above. 
So, it is expedient to search for the other means
to make precision measurements in a wide region of possible Higgs boson mass 
values. In this respect a crucial role will be played by  colliding lepton 
beams.  Among them, in turn, the future $\mu^+ \mu^-$ colliders  will be more 
prefereable in this respect than electron-positron one as it will 
be seen below.
    
    For the first time an idea of colliding $\mu$ -- meson beams    
together with its physics potential was discussed in paper \cite{3}.
Later this question was raised in refs. \cite{4}, \cite{5}, \cite{6}.
At present concrete proposals and the technical designs for the 
$\mu^+ \mu^-$ colliders are being intensively discussed together with 
the physics goals (see, e.g. ref.\cite{7}).
  Up to now  there are two designs for the future $\mu^+\mu^-$--colliders:
\begin{enumerate}
\item  [1)] $\sqrt{s} \,\approx \, 500$ GeV, ${\cal L} \,\approx \, 10^{33}$
            cm$^{-2}$ s$^{-1}$, $L_{tot}\,=\,50$ fb$^{-1}$/year; 
\item  [2)] $\sqrt{s} \,\approx \, 4$ TeV, ${\cal L} \,\approx \, 10^{35}$
            cm$^{-2}$ s$^{-1}$, $L_{tot}\,=\,200 \,-\,1000$ fb$^{-1}$/year.
\end{enumerate} 

    A muon collider has some natural advantages as compared to an $e^+e^-$
collider, including some that are important for Higgs bosons production 
\cite{7}, \cite{8}:
\begin{enumerate}
\item [1)] there is essencially no beamstrahlung;
\item [2)] there is substantially no bremstrahlung;
\item [3)] $\mu^+\mu^-$ collider has a higher mass of initial particles
in comparison with $e^+e^-$ collider;
\item [4)] there is no final focus problem (storage rings are used to build up
the effective instantaneous luminosity);
\item [5)] rather high beam energy resolution of $R\approx 0.1 \%$
is possible, if the necessary technology is built into the machine;
\item [6)] region $\sqrt{s_{max}}\, \geq \,500 $ GeV can probably be 
reached more easily;
\item [7)] much less storage ring diameters are required because of
drastic reduction of synchrotron radiation, correspondingly, cost of this 
part of design is reduced substantialy.
\end{enumerate}

    The negatives regarding a $\mu^+\mu^-$ collider include \cite{7}, \cite{8}:
\begin{enumerate}
\item [1)] the design is immature, and approximately five years of research and
development projects are needed before a full--fledged proposal would be 
possible -- in particular, cooling tests are required to see if multistage
cooling will be sufficiently efficient;
\item [2)] the exact nature of the detector backgrounds, and how to manage
them, is still under investigation -- certainly the detector will be more
expensive due to higher shielding requirement;
\item [3)] significant polarization probably implies significant loss in
${\cal L}$;
\item [4)] it is not possible to have $\gamma\gamma / \gamma \mu$ facilities;
\item [5)] due to the $\mu^+\mu^-$ initial state, we have mainly $J\,=\,1$
in comparison with almost all angular momenta for $\gamma\gamma / \gamma \mu$
facilities;
\item [6)] in many important cases the cross--sections behaviour is
$\sigma\,\approx\,1/s$ for $\mu^+\mu^-$ collisions and
$\sigma\,\approx\,const$ for $\gamma\gamma / \gamma \mu$ facilities;
\end{enumerate} 

    Almost all theoretical efforts lie in the threshold $s$--channel 
SM Higgs boson production as depicted on Fig. 1.
    But in parallel with some advantages (such as threshold behaviour),
there are also a number of problems:
\begin{enumerate}
\item [a)] $\mu^+\mu^-\,\to\, b\bar b$ process requires extreme beam energy
resolution, e.g.  of the order of $R\, \approx\, 0.01 \%$ or higher. 
Inspite of great possibilities of the $\mu^+ \mu^-$ colliders in this domain, 
that may be unattainable;
\item [b)] the mass of the SM Higgs boson must be a priory reported from 
another sources, e.g. LHC measurements. This also may have a problem. 
The precision of the 
LHC measures supposed is of the order  $\delta m_H \approx \, 1 \% \cdot m_H$ 
in $\gamma\gamma$ decay mode \cite{7}. For $m_H\,\approx\, 200 \,-\,300$ GeV, 
the error on the Higgs boson mass is about $\delta m_H\,\approx\, 2\,-\,3$ GeV, 
and we have rather broad range for scanning;
\item [c)] for precision measurements too high luminosity ${\cal L}$ must be
achieved at all the scanned energies;
\item [d)] in any case, the several final storage rings designed to maintain
near--optimal ${\cal L}$ over a span of $\sqrt{s}$ values are to be
constructed.
\end{enumerate} 

   In view of this  it would be very interesting, if there exist
other processes with SM Higgs boson production, which haven't all or part of 
the above--listed disadvantages.

    One of such processes is a reaction $\mu^+ \mu^- \, \to \, H^0 \gamma$ 
(see Fig.2). 

    In the $e^+ e^-$ collisions the contribution of those  diagrams to the
overall cross--section is extremely small in comparison with higher order 
diagrams with heavy particles in loops \cite{9}. In the case of $\mu^+ \mu^-$
collisions the lowest order diagrams are competetive with loop diagrams
due to a greater mass of $\mu$ in comparison with the electron mass. That 
process may be one of the goals for the future $\mu^+ \mu^-$ colliders. 

   Yet, a  related process exists, which may be be even more interesting,
namely $\mu^+ \mu^- \, \to \, Z H^0$. The corresponding Feynman graphs
are depicted on Fig.6. This process differs from the above in
that its cross-section is not negligible at tree level due to 
the s--channel diagram, Fig.6-c. The contribution of the remaining two 
graphs, Fig.6a-b, to the cross-section is negligible for the case of
$e^+ e^-$ collisions, however, in the case of $\mu^+ \mu^-$ collisions
their contribution is finite. Moreover, only due to accounting for  
them it is possible to obtain the correct asymptotic behaviour of cross
section, when initial particles masses are involved into calculation.
This phenomenon reflects one of the fundamental property of the theory of
electroweak interaction \cite{10}.  This question will be thoroughly
discussed in the section 3. 

    The rest of the paper is organized as follows. In Section 2 we
analyse the associated Higgs boson -- photon production in the Standard 
Model. In Section 3 we investigated the prospects for the associated 
Higgs--Z-boson production. Section 4 contains Conclusions.

\section{Associated {\boldmath $H^0\gamma$} production in SM}

    In the Standard Model the process $\mu^+ \mu^- \, \to \, H^0 \gamma$
is described to lowest order by the Feynman diagrams, depicted in Fig. 2.

    Summing over the polarizations of the photon and averaging over the
polarization of both the initials $\mu^+ \mu^-$ beams, the differential
cross--section of process (1) can be written as:

\begin{eqnarray} 
\frac {d\sigma}{d cos\theta}(\mu^+\mu^-\,\to\,H^0 \gamma )\,=\,
\frac{\pi\alpha^2}{8sin^2\theta_W}\cdot\frac{m^2_{\mu}}{M^2_W}\cdot 
\frac {1}{s^2}\cdot\frac{1}{\beta}\cdot (s\, -\, m^2_H) \times \nonumber \\
\Biggl\{\frac{1}{(k \cdot p_1)^2}\biggl[(k \cdot p_1)(k \cdot p_2)
~+~ m^2_{\mu}\cdot\bigl[-(k \cdot p_1) + 
(k \cdot p_2)-\frac{1}{2}s\beta^2 \bigr]\biggr] \nonumber \\
~+~\frac{1}{(k \cdot p_2)^2}\biggl[(k \cdot p_1)(k \cdot p_2)
~+~ m^2_{\mu}\cdot\bigl[(k \cdot p_1) -
(k \cdot p_2)-\frac{1}{2}s\beta^2 \bigr]\biggr] \nonumber \\
~+~\frac{2}{(k \cdot p_1)(k \cdot p_2)}\biggl[ (k \cdot p_1)(k \cdot p_2)
- \frac{1}{2}s\beta^2\cdot(-\frac{1}{2}m_H^2+m^2_{\mu})\biggr]   \Biggr\}
\end{eqnarray} 
where the following notations were introduced
\begin{eqnarray} 
 (k \cdot p_{1,2})\,&=&\,\frac{1}{4}(s-m_H^2)(1\mp\beta cos\theta), \nonumber \\
\beta~&=&~\sqrt{1-\frac{4m^2_{\mu}}{s}}
\end{eqnarray} 
with $\sqrt{s}$ as the c.m. energy and $\theta$ -- the scattering angle of
the photon. 
After introducing in addition to the usual $\beta$ the notation
\begin{eqnarray} 
\beta_H~=~\sqrt{1-\frac{4m^2_{\mu}}{m_H}}
\end{eqnarray} 
and integration over $cos \theta$ in the $[-1,1]$ limits, the 
cross--section acquires the following final form:

\begin{eqnarray} 
\sigma(\mu^+\mu^-\,\to\, H^0\gamma) \,=\,
\frac{\pi\alpha^2}{2\sin^2\theta_W}\cdot\frac{m^2_{\mu}}{M^2_W}\cdot
\frac {1}{s^2} \cdot \frac{1}{ \beta}\cdot
\frac{1}{s\,-\,m^2_H}\times \nonumber \\
\Biggl\{-2m^2_H\,s\,\beta_H^2\,+\,\bigl(s^2\,\beta^4+\,m^4_H\,\beta_H^2\,
\bigr)\cdot\frac{1}{ \beta}\cdot
\ln\frac{1\,+\, \beta }
{1\,- \, \beta }\Biggr\}
\end{eqnarray} 

    All the calculations have been performed with nonzero muon mass.

    The cross--sections for the process $\mu^+\mu^-\,\to\, H^0\gamma$ are shown
in Fig.3 as a function of the Higgs boson mass for the three center of mass 
energies: $\sqrt{s}\,=\,500$ GeV, $\sqrt{s}\,=\,1$ TeV,$\sqrt{s}\,=\,4$ TeV.
At 500 GeV the cross--section is of the order of $\sigma\,=\,2 \div 3 \times
10^{-2} fb$ for the light Higgs boson masses. At $\sqrt{s}\,=\,1.5$ TeV 
the cross--section for light Higgs boson drops by a factor of $\,\approx 4$ 
compared to the previous case. 

    Fig.4 exhibits the dependence of the cross--section on the center of mass
energy for several values of the Higgs boson mass. 
The cross section decreases smoothly with the increasing energy; it scales 
approximately as $ln s/s$ at high energies.

    The asymptotic behaviour of the cross--section under condition
$\sqrt{s}\,\to\,\infty$ and $s\,\gg\,m^2_H$ can be written as:
\begin{eqnarray} 
\sigma_{as}(\mu^+\mu^-\,\to\,H^0 \gamma)\,=\,
\frac{\pi\alpha^2}{2\sin^2\theta_W}\cdot\frac{m^2_{\mu}}{M^2_W}\cdot
\frac{\ln\bigl(s/m_{\mu}^2\bigr)}{s}
\end{eqnarray} 

    Finally, Fig.5 shows the angular distribution $d\sigma/d cos \theta$ 
for several Higgs boson mass values at c.m. energy of 500 GeV. 
The distribution is forward--backward
symmetric and does not depend very strongly on the Higgs boson mass.

    With the yearly integrated luminosity of ${\cal L}\, \approx \, 10^3$
fb$^{-1}$ expected at future $\mu^+ \mu^-$ colliders, one could collect 
20 to 30 $H^0\gamma$ events (detector efficiency is supposed equal 1, and
acceptance -- $4\pi$). The signal, which mainly consists of a photon and 
$b \bar b$ pairs in the low Higgs mass range or $WW/ZZ$ pairs for Higgs masses
larger than $\approx \, 200$ GeV, is extremely a clean. The backgrounds 
should be very small since the photon must be very energetic and the 
$b \bar b$ or $WW/ZZ$ pairs should peak at an invariant mass $M_H$. 
Therefore, despite of the low rates, a clean signal gives a good 
possibility to detect these events.

    Expressions  (2) -- (4) obtained for the cross-section of the
process $\mu^+ \mu^- \, \to \, H^0 \gamma$ are applicable, on the equal
foot, to the case of any other (pseudo)scalar particles production. 
Moreover it might happen that namely muon colliders will be most suitable
and crucial means for their searches. Foremost it refers
to the axion. This particle was postulated in papers \cite{11} and \cite{12} 
as a consequence of the strong $CP$ -- violation  problem solution \cite{13},
\cite{14}. The numerous fruitless searches of that pseudoscalar 
(for review see, e.g. ref. \cite{15}) produced a widely accepted opinion, 
that this hypothetical particle is extremely light and weakly interacting 
one ("invisible axion", \cite{16}). 

      However in a recent paper \cite{17} 
the solution of strong $CP$ -- violation problem in $QCD$ has been proposed, 
which may lead to a heavy axion, $M_a\,\leq\,1$ TeV. Its interaction with 
usual matter is induced by mixing with axial Higgs boson. For example, 
in the case of fermions it has the form $ {\bf {\cal L}_{int} \sim const \cdot 
m_f \cdot (a\tilde{f}i\gamma_5f)}$. A mixing parameters are model dependent 
but might not be negligible small, therefore this interaction can lead to 
a observable effects. In that case the muon colliders might be 
irreplaceable tool for the axion search aim.

\section{Associated {\boldmath $HZ$} production in SM}

    Another interesting process for the Higgs investigations is  
$\mu^+ \mu^- \, \to \, Z H^0$.  At first sight it is analogous to the process
considered in the preceding section. However, it possesses the additional
very interesting features, which display the deepest properties of the
nonabelian gauge theories with spontaneous symmetry breaking. First of all,
one finds the difference in the numbers of Feynman graphs, corresponding to
both of aforementioned processes. For the second of them they are 
drawn on Fig.6. To the third diagram of this set, Fig.6-c, corresponds
the so called Bjorken process, considered early for the case of 
$e^+ ~ e^-$ collisions \cite{18}( see also \cite{19}). All those calculations 
had been done in the limit where the masses of initial particles were
neglected. Now, with the accounting for those masses the cross-section reveals
a very interesting feature: despite of its $s$--channel character it does not
fall at very high energy but approaches a constant limit. At the same time
its angle distribution is flat, indicating that it comes entirely from
the $J\,=\,0$ plane wave. It is obvious, that this behaviour contradicts 
unitarity condition, which requires $\sigma_{J=0} ~\leq~s^{-1}$ at high
energy. The contradiction is removed if in calculation procedure all the
three diagrams of Fig.6 are accounted for. Because the whole contribution
of $t$--channel diagrams of Fig.6 is proportional to the initial 
particles mass it might be considered as an additional argument in favor 
of the $\mu$--meson collider.

     In the course of cross-section calculation for the process 
$\mu^+ \mu^- \, \to \, Z H^0$  without neglecting the masses of initial 
$\mu^\pm$ much more complicated expressions arise so we confine 
ourselves by numerical computation with the aid of the Monte Carlo 
method for integration on phase space of final particles to obtain
the total cross--section and various distributions.

    The main formula for the Monte Carlo calculations is
\begin{eqnarray} 
\sigma\,=\,\int \, f(\stackrel{\rightarrow}{\Phi})\,
d\stackrel{\rightarrow}{\Phi}
\end{eqnarray} 
where $f(\stackrel{\rightarrow}{\Phi})$ denotes the matrix element squared
(any cut can be easily implemented by putting $f(\stackrel{\rightarrow}{\Phi})
\,=\,0$ in the unwanted region of the phase space), and $d\stackrel
{\rightarrow}{\Phi}$ is the 2--body phase space integration element.

    In view of vital importance of remarks, made in the beginning of
this section it is expedient to discuss separately contributions to 
the cross-section of the first two diagrams of Fig.6 from the one hand 
side and the third one from the another hand side. Fig.7 (lower curve) shows 
the c.m. energy dependence of contribution to the cross-section of the
sum of the first two diagrams, Fig.6a-b, along with the contribution of 
whole set of diagrams (upper curve). Already at the relatively not too high 
energy the contribution of t -- channel  diagram plus u -- channel one
approaches the limiting value equal to $\approx 1.2~\cdot~10^{-2}~fb$. 
     
      Now, let us calculate the cross-section corresponding to the 
diagram Fig.6-c alone, accounting for masses of initial muons. 
Asymptotics of this process at $\sqrt s \rightarrow \infty$ is as follows:

\begin{eqnarray} 
\sigma^{(c)}_{as}(\mu^+ \mu^- \, \to \, Z H^0)|_{m_{\mu}\neq0}
 ~=~ \frac{2\pi~\alpha^2}
{\sin^4(2\theta_W)}\cdot g^2_A \cdot\frac{m^2_\mu}{m^4_Z}
\end{eqnarray} 
 
It is seen, that despite of the fact that this diagram is the pure $s$--channel 
one, the corresponding cross-section is not falling at high energy, but
approaches a constant limit, whose value is  also equal to $\approx 1.2 ~
\cdot~10^{-2}~fb$. There is sence to draw attention to the absence of vector 
coupling in the expression obtained. At last, the interference term between sum 
of t -- and u -- channel diagrams of Fig.6 and those of s -- channel gives 
the contribution to the full cross-section, which is equal to 
$\- approx 2.4 ~\cdot~10^{-2}~fb$. Therefore, we see that only accounting 
for all of three diagrams on Fig.6 with finite muon mass gives the correct 
asymptotic behaviour of cross section.

Note, that usually the calculations were $e^+~ e^-$ oriented with electron
mass neglected, so the Yang-Mills character of theory was enough to
secure the situation. In our case the Higgs mechanism is urgently needed. 
Concerning the process considered the tuning compensation would allow for
studying a new physics or to feel the existence of more complicated
Higgs sector. Evidently muon colliders will deliver a unique 
possibility to study interactions of Higgs scalar within the lepton sector.

  Fig.8 shows the angular distribution $d\sigma/d cos \theta$ for
c.m. energy of 500 GeV, 1 TeV, 4 TeV and for a Higgs boson mass $M_H\,=\,
100$ GeV in all the three cases. The distribution is forward--backward
symmetric and does not depend very strongly on the Higgs boson mass.
It reveals a typical behaviour of the scalar particle emitted when
fermion-antifermion pair collide  and fuse into vectorial one (Z -- boson
in the case at hand), i.e. it prefers to fly at $90^0$. Explicitly,
the corresponding piece of differential cross-section behaves as 
$\approx a\,-\, b\cdot\cos\theta$ with $a$ and $b$ being positive.

    Fig.9 exhibits the dependence of the cross--section on the center of mass
energy for several values of the Higgs boson mass. The cross--section increases
rapidly with the opening of the phase space and then drops
smoothly with the increasing energy; it scales approximately as $1/s$ at
high energies. Explicitly, the asymptotic behaviour for $\sqrt{s}\,\to\, 
\infty$ of the cross--section is as follows:
\begin{eqnarray} 
\sigma^{as}(\mu^+\mu^-\,\to\,H^0 Z)
\,=\,\frac{1}{3}\cdot
\frac{\pi\alpha^2}{sin^4(2\theta_W)}\cdot\Bigl(g^2_V\,+\,g^2_A\Bigr)
\cdot\frac{1}{s}
\end{eqnarray}

    The cross--sections for the process $\mu^+\mu^-\,\to\, H^0 Z$ are shown
in Fig.10 as a function of the Higgs boson mass value for three representative
center of mass energy, $\sqrt{s}\,=\,500$ GeV, $\sqrt{s}\,=\,1$ TeV,
and $\sqrt{s}\,=\,4$ TeV.    At 500 GeV the cross--section is of 
the order of $\sigma\,=\,10^{-1}pb$ for the light Higgs masses; it 
drops out slightly with increasing $M_H$ due to the 
lack of the phase space. At $\sqrt{s}\,=\,1$ TeV the cross--section 
for light Higgs boson drops by a factor of $\approx 9$ compared to the previous
case, but the decrease with increasing $M_H$ is slower.

    With the yearly integrated luminosity of ${\cal L}\, \approx \, 10^3$
fb$^{-1}$ expected at future $\mu^+ \mu^-$ colliders, one could collect a
sufficient number of $H^0 Z$ events for thorough investigations of this
process (detector efficiency is supposed equal to 1, and acceptance -- 
$4\pi$). The signal, which mainly consists of a Z--boson products
and $b \bar b$ pairs in the low Higgs mass range or $WW/ZZ$ pairs for 
Higgs masses larger than $\approx \, 200$ GeV, is rather clean. 
The backgrounds should be rather small since the Z--boson must be very 
energetic and the $b \bar b$ or $WW/ZZ$ pairs should peak at an invariant 
mass $M_H$. Therefore, the clean signal gives a good possibility for
extensive study of these events.

\section{Conclusions}

    We have calculated the cross--sections for the production of the
Standard Model Higgs boson in association with a photon and Z--boson in 
$\mu^+ \mu^-$ collisions in the lowest order of the perturbation theory. 
We have given the complete and compact analytical expression for 
$\mu^+ \mu^-\,\to\,H^0\gamma$ process with detailed Monte Carlo simulations. 
For the case of $\mu^+ \mu^-\,\to\,H^0 Z$ process, we presented analytically 
only asymptotyc expressions for the cross--section of the process; 
all the histograms were produced by means of Monte Carlo simulation.

    We have then illustrated the size of the cross--sections for energies,
which will be reached at future $\mu^+\mu^-$ colliders. The cross--section 
for $\mu^+ \mu^-\, \to\,H^0\gamma$ process is, in general, small, but much 
more intensive compaired with the corresponding signal for the case of 
$e^+ e^-$ collisions (at tree level), and rather clean. With an integrated 
yearly  luminosity of ${\cal L}\, \approx\, 1000$ fb$^{-1}$ expected at 
future $\mu^+\mu^-$ colliders, we can detect those signals despite 
the low rates. Process $\mu^+ \mu^-\,\to\,H^0 Z$, in turn, is easly 
detectable and gives some opportunity to study the Higgs boson interaction
in the lepton sector. From the theoretical point of view it demonstrates
the efficiency of modern electroweak interaction scheme.

    In parallel with threshold $s$--channel Higgs boson production, the
processes $\mu^+\mu^-\,\to\,H^0\gamma$ and $\mu^+\mu^-\,\to\,H^0 Z$ 
may be used for the search for mass of the SM Higgs boson.

\section{Acknowledgments}
      We are grateful to Dr. V.I.Balbekov for clarifying some
questions concerning the accelerator technique. One of us (F.F.T)
is indebted to S.R.Slabospitsky for the help in drawing the Feynman
graphs, Fig.1, Fig.2, Fig.6, with the aid of DURER macropackage.
\newpage

\centerline{\bf Figures captions}
\begin{enumerate} 
\item [ Fig.1.] Diagram for $s$--channel Higgs boson production is shown.
\item [ Fig.2.] Diagrams for $\mu^+\mu^-\,\to\,H^0\gamma$ are shown.
\item [ Fig.3.] The cross section for $\mu^+\mu^-\,\to\,H^0\gamma$ is
                given as a function of Higgs boson mass 
                for collider energies of $\sqrt{s} \,=\,500$ GeV,
                $\sqrt{s}\,=\,1$ TeV and  $\sqrt{s}\,=\,4$ TeV.
\item [ Fig.4.] The cross section for $\mu^+\mu^-\,\to\,H^0\gamma$
                is given for several values of Higgs boson mass. Curves
                correspond to $M_H$ = 100, $M_H$ = 150, and  $M_H$ = 200 GeV.
\item [ Fig.5.] The angular distributions $d\sigma/d cos \theta$ for
                $\mu^+\mu^-\,\to\,H^0\gamma$ are shown for 
                collider energies of 500 GeV and several values of $M_H$.
\item [ Fig.6.] Diagrams for $\mu^+ \mu^- \, \to\, Z H^0$ are shown.
\item [ Fig.7.] The  cross-section for $\mu^+ \mu^- \, \to\, Z H^0$
                resulting from the sum of the diagrams drawn in Fig.6a-b 
                is shown (lower curve) along with the cross-section 
                resulting from full set of diagrams drawn in Fig.6
                (shown at upper curve).
\item [ Fig.8.] The angular distributions for $\mu^+ \mu^- \, \to\, Z H^0$
                are shown for collider energies of 500 GeV, 1 TeV and 4 TeV.
                $M_H$ = 100 GeV in all the three cases.
\item [ Fig.9.] The cross--section for $\mu^+ \mu^- \, \to\, Z H^0$
                is given for several values of the Higgs boson mass,
                $m_H\,=\, 100, \,150 and \,200$ GeV
\item [ Fig.10.] The cross--section for $\mu^+ \mu^- \, \to\, Z H^0$
                 is given as a function of Higgs boson mass for 
                 collider energies of 500 GeV, 1 TeV and 4 TeV.
\end{enumerate}

\bigskip
\bigskip
\bigskip
~~~~~~~~~~~~~~~~~~~~~~~~~~~~~~~~~~~~~~~~~~~~~~~~~~~~~~~~~~~{\it Received 
April 17, 1997}

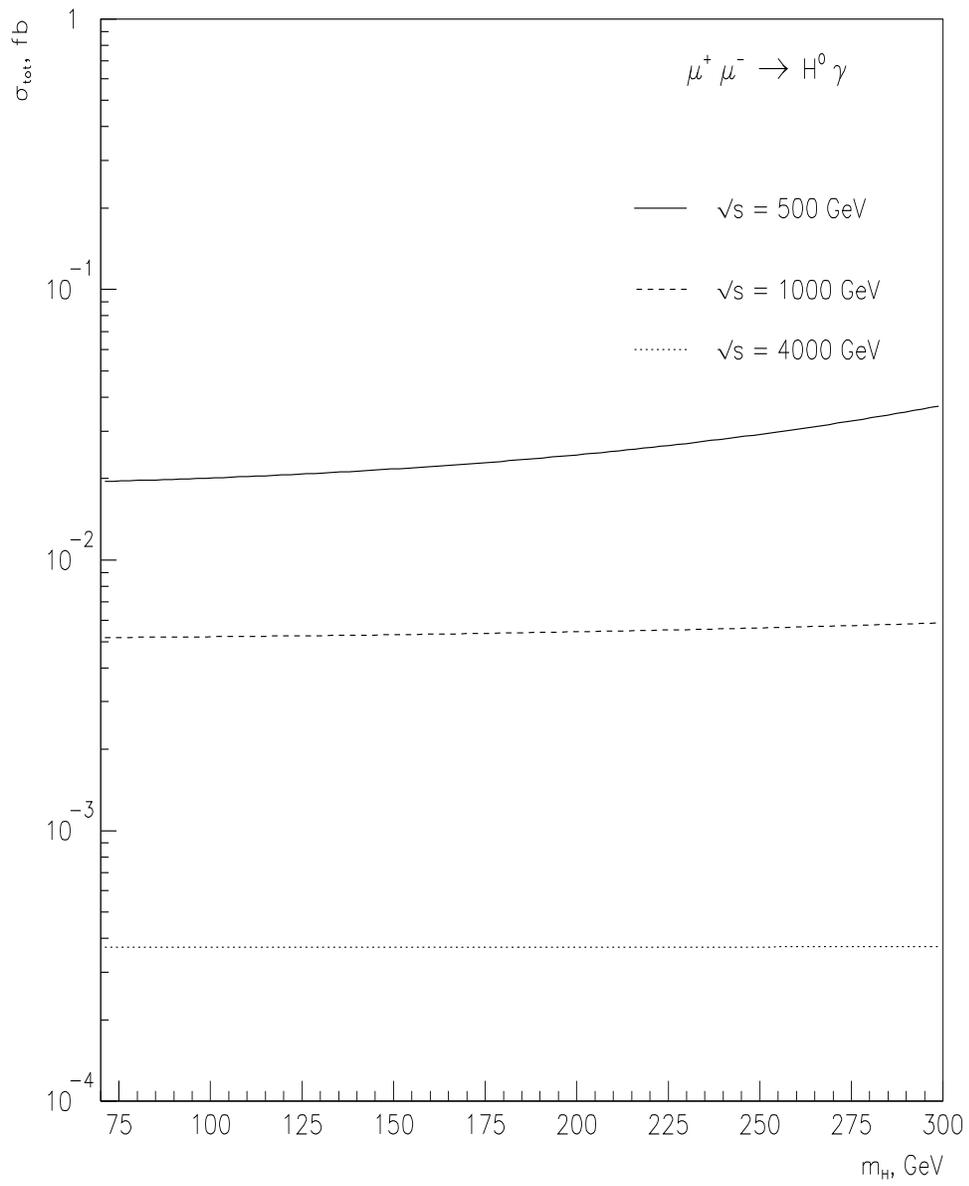
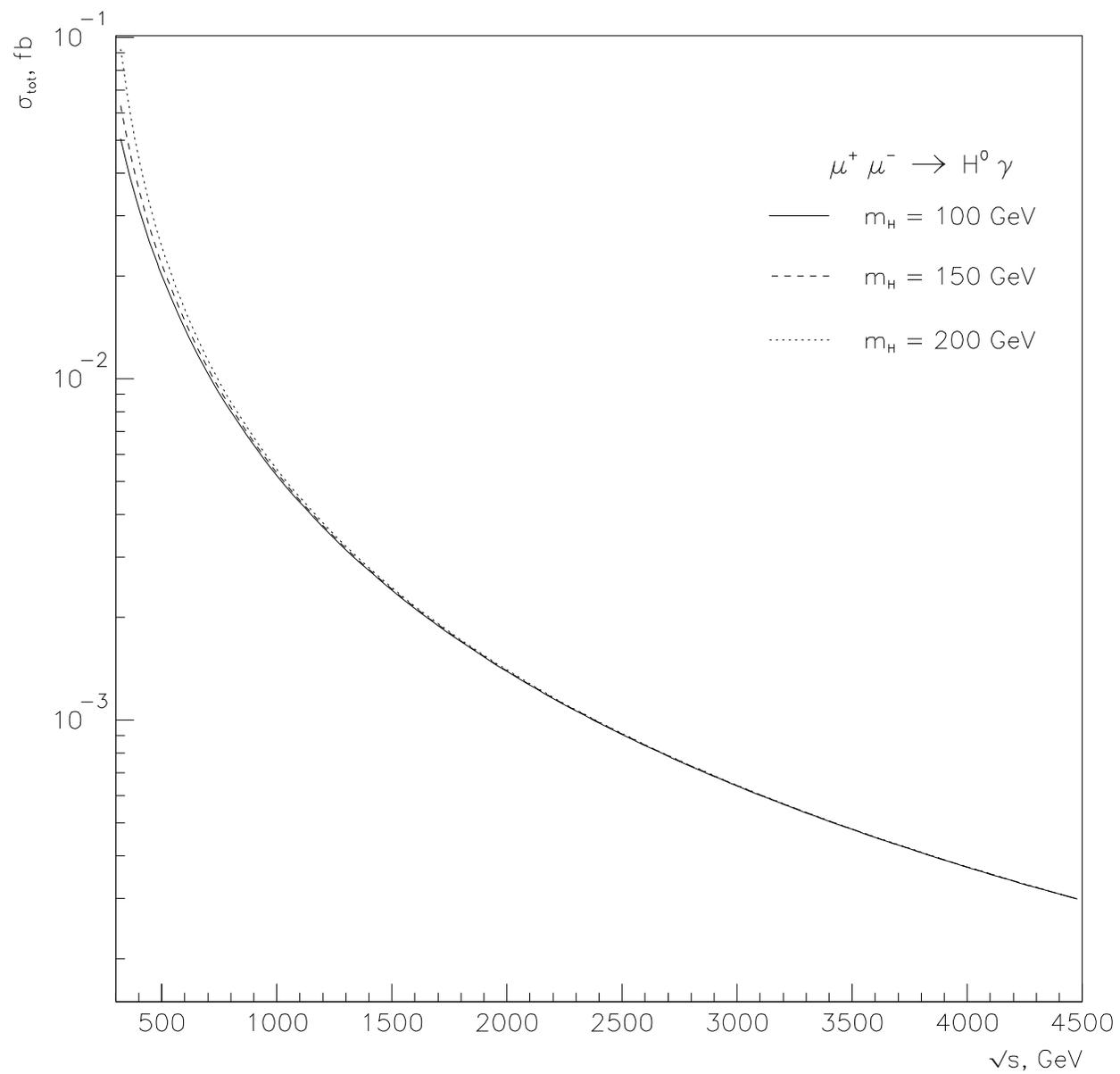
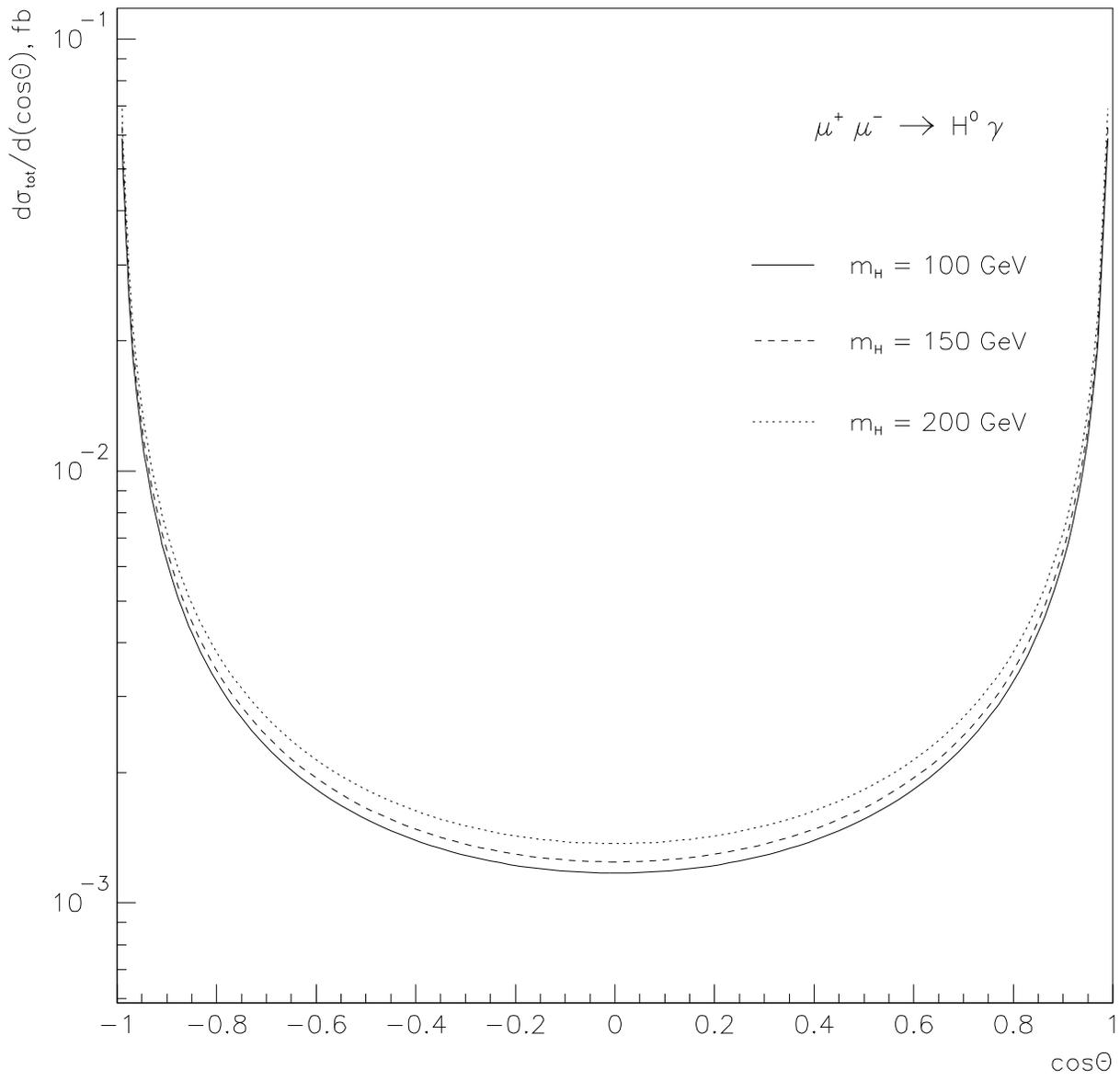
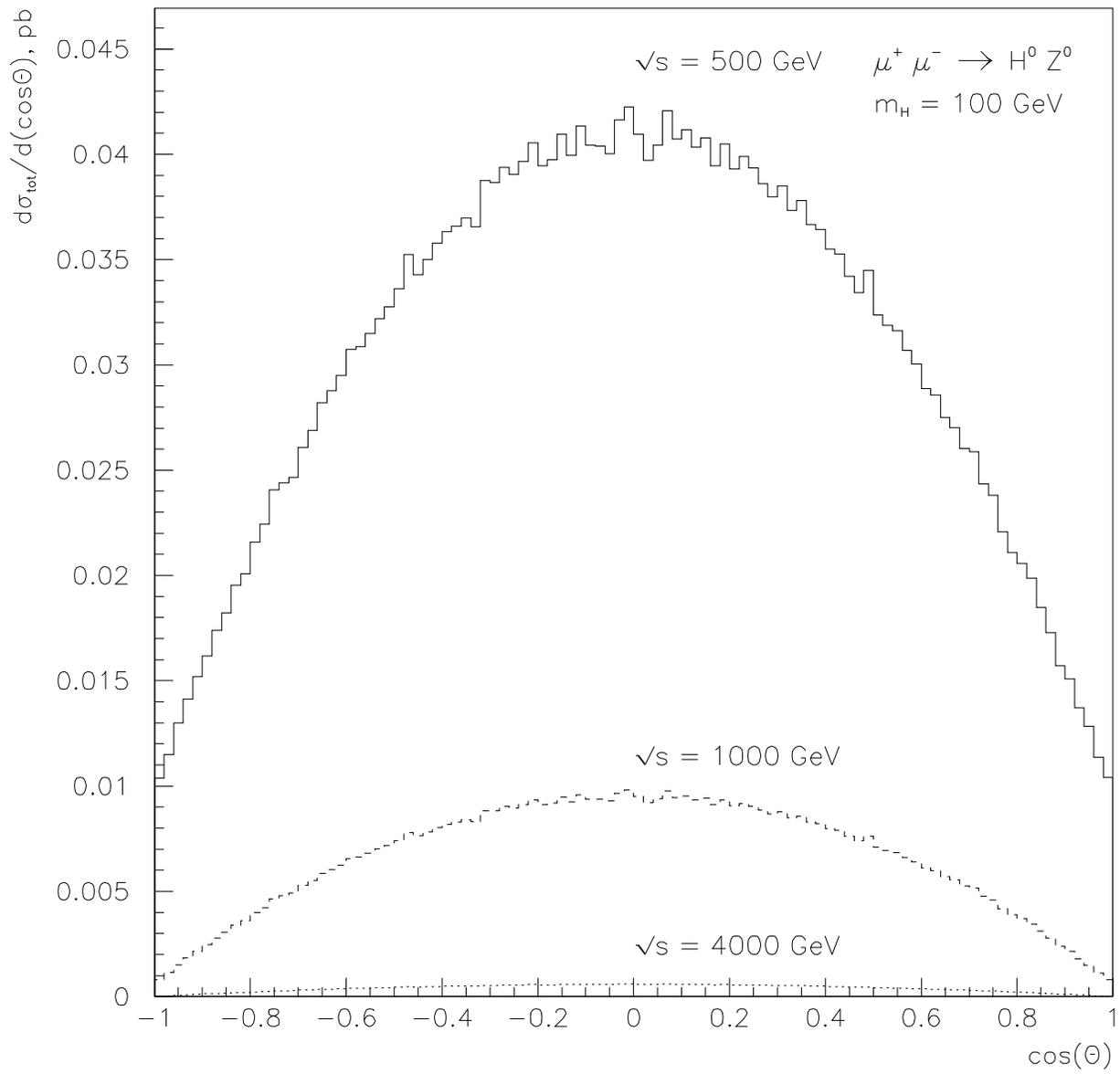
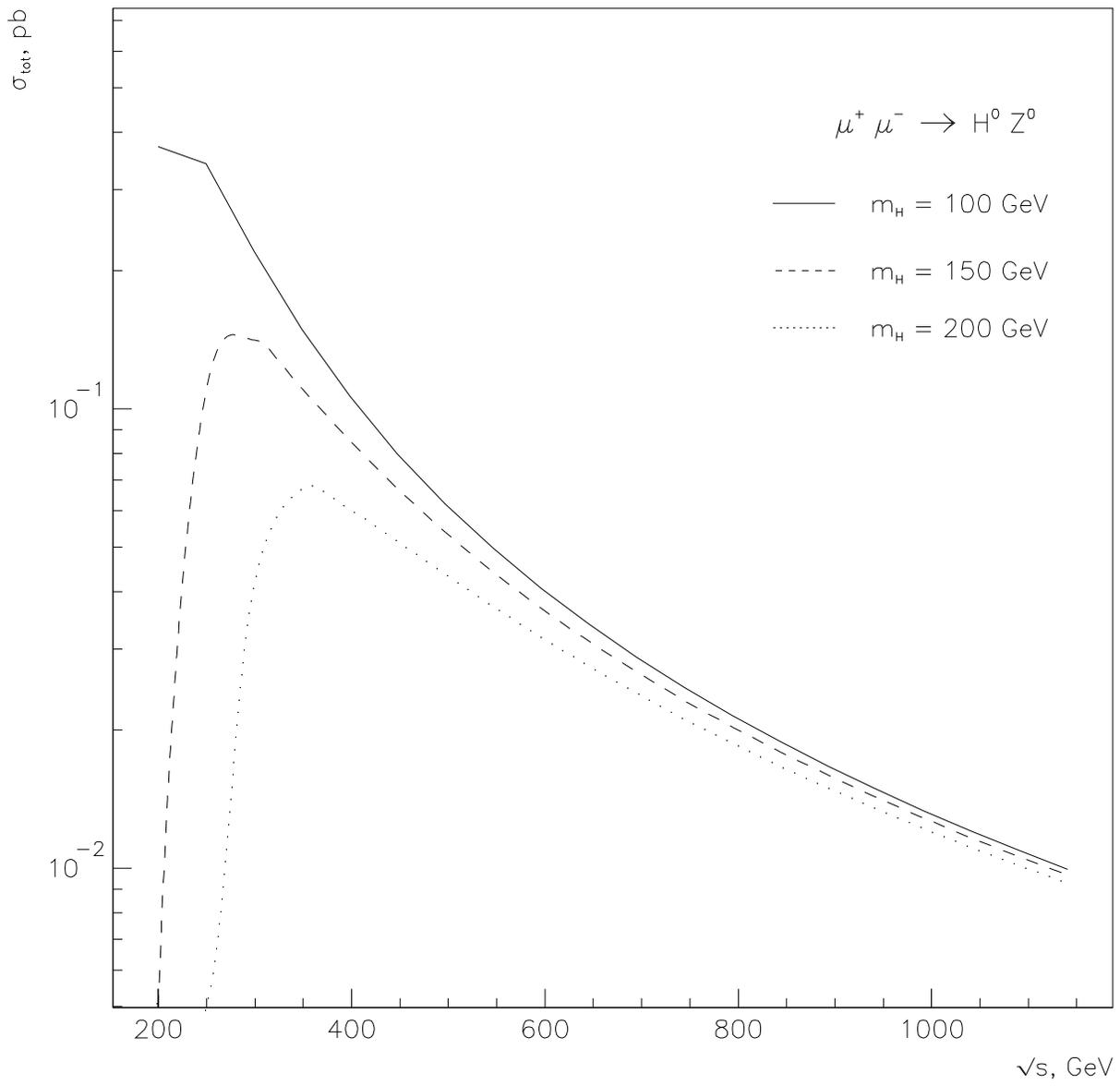
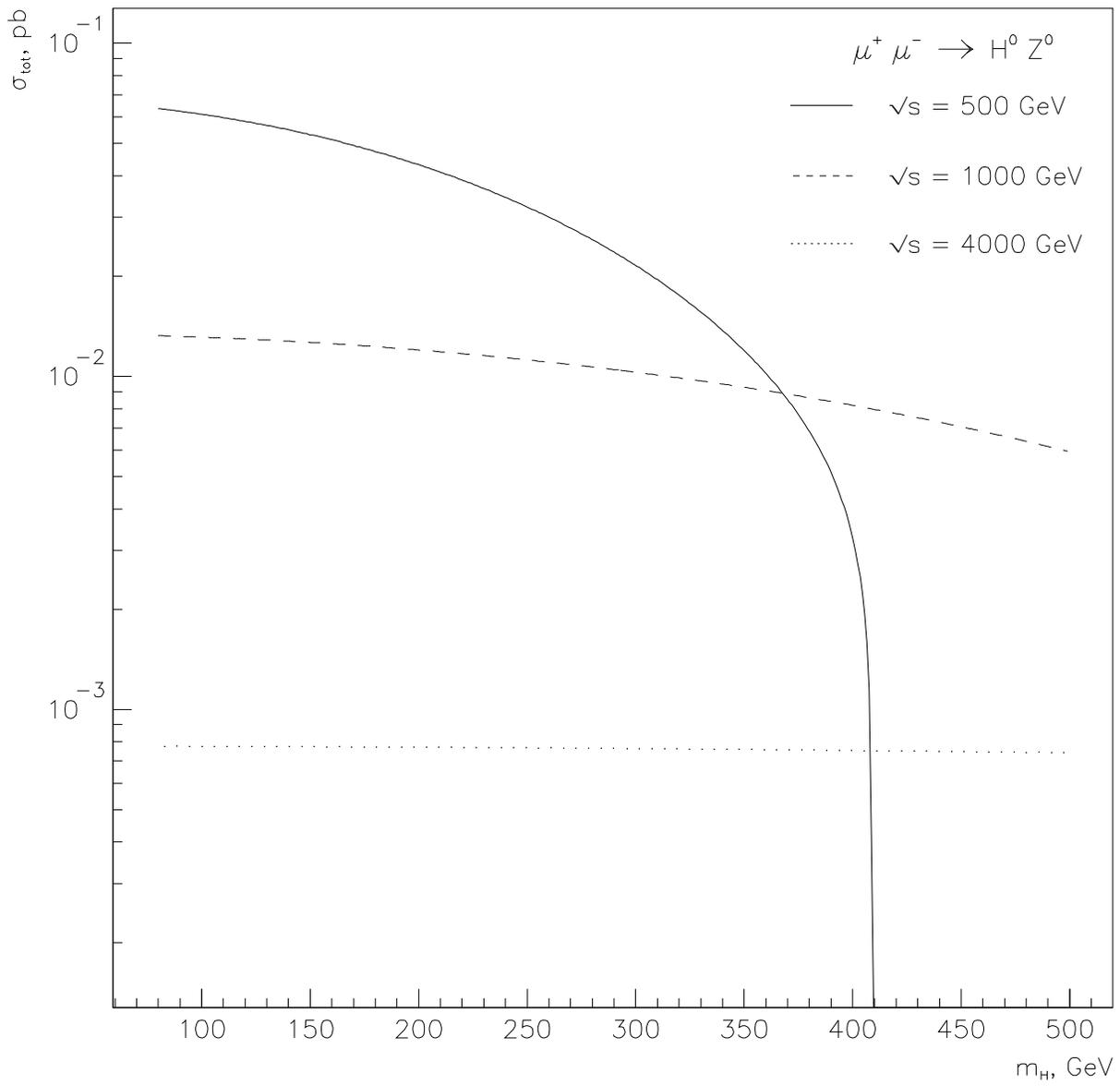
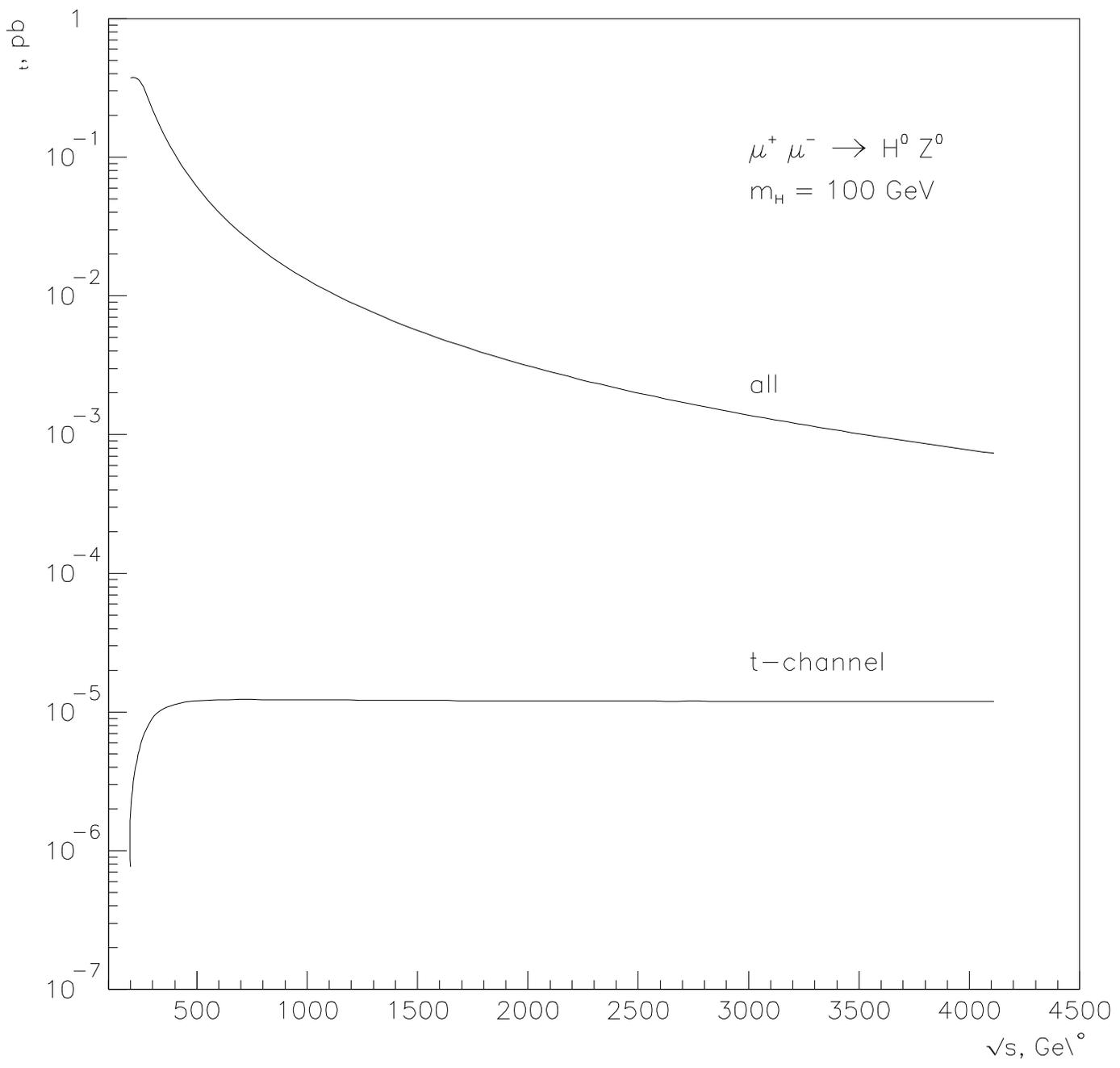
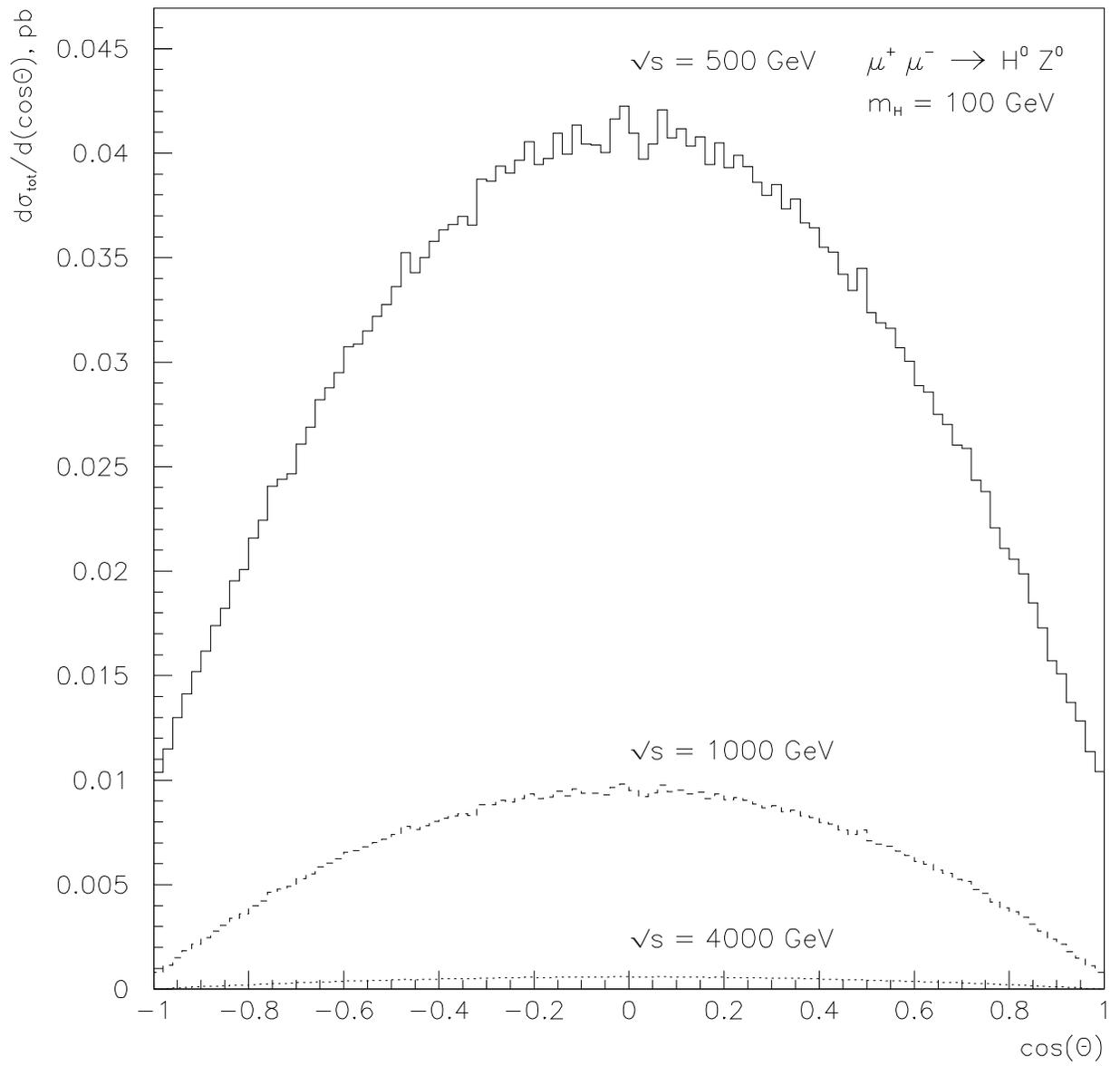
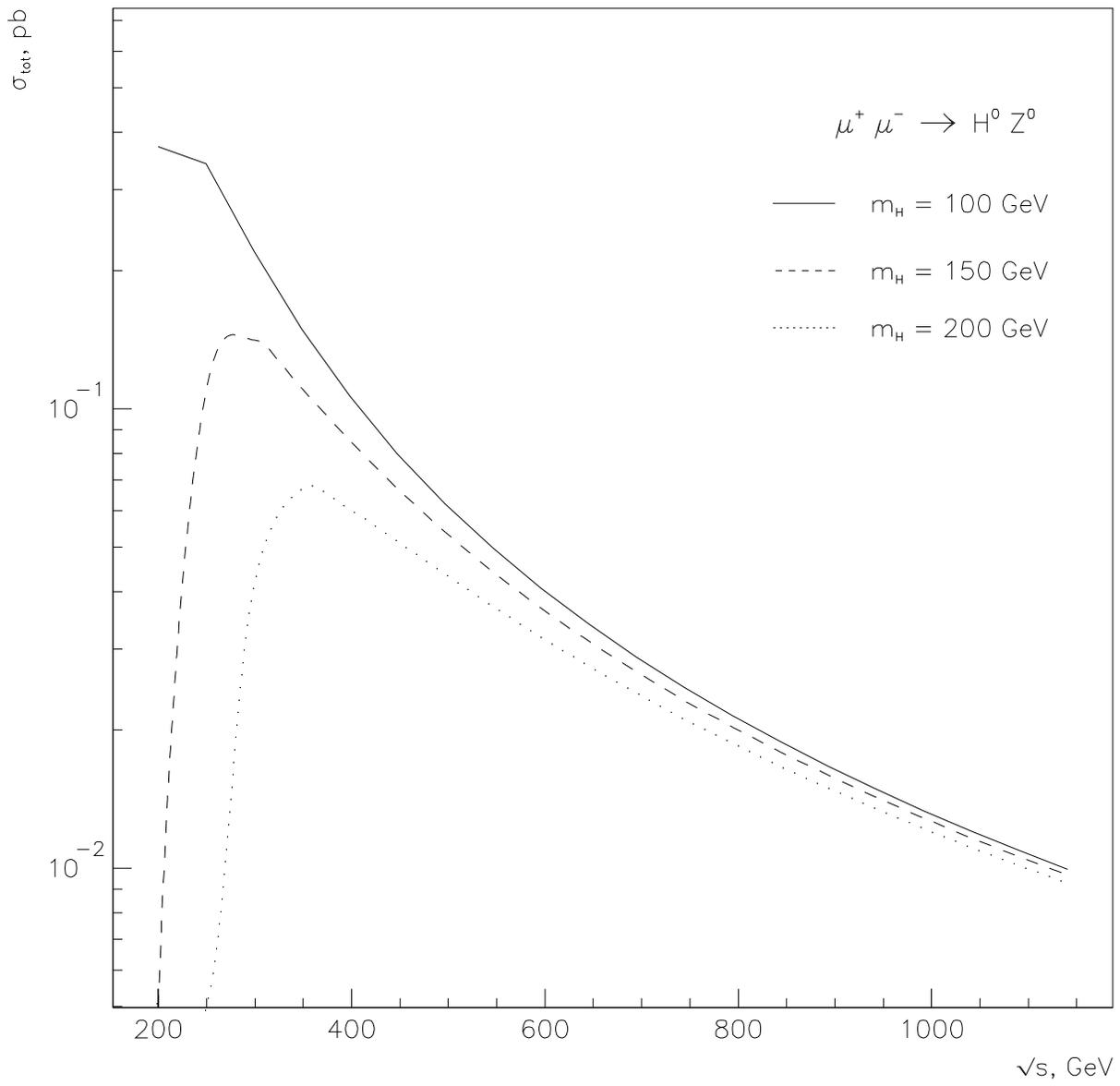
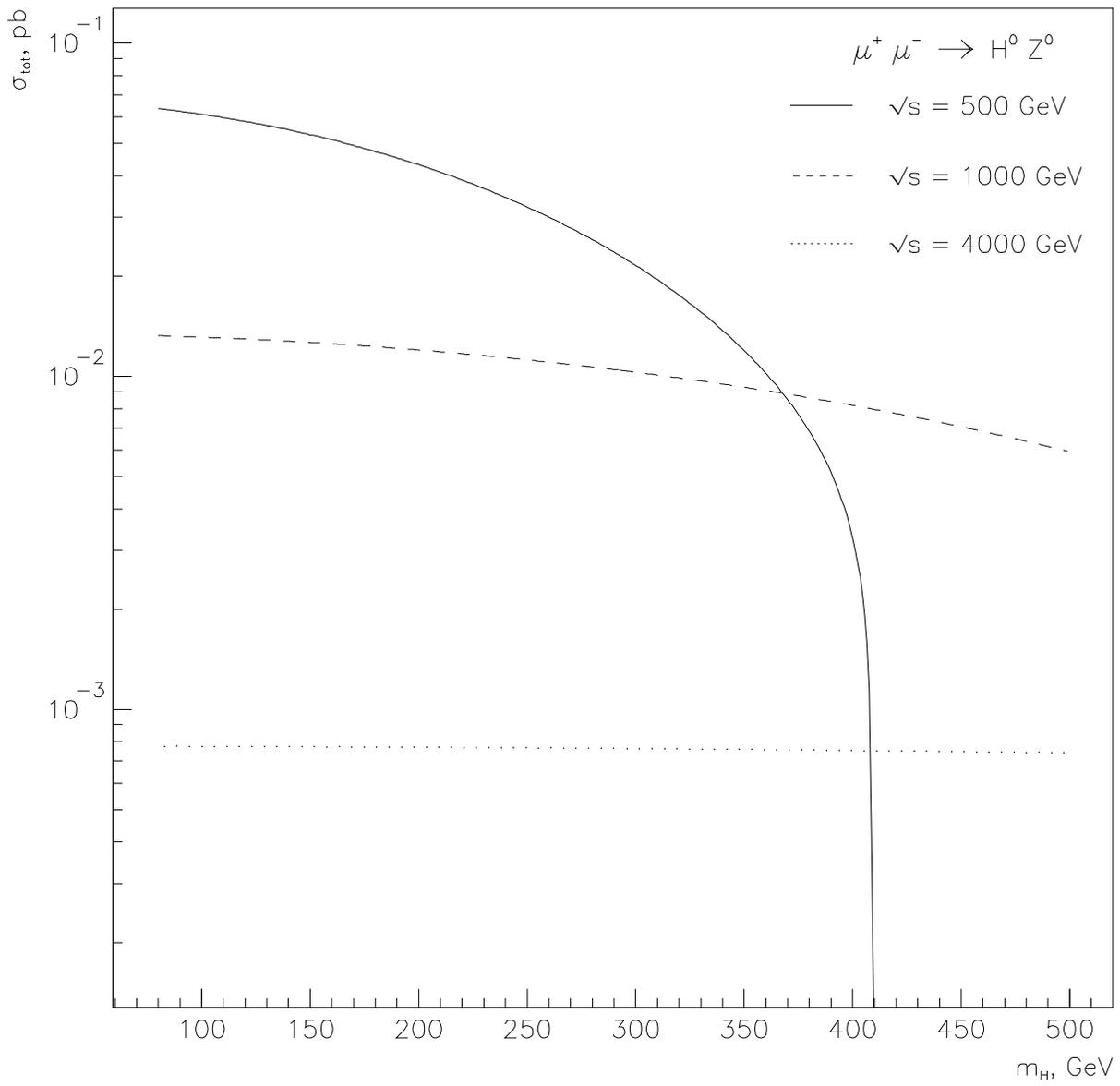
\end{document}